\documentclass{article}

\usepackage{arxiv}

\usepackage[utf8]{inputenc} % allow utf-8 input
\usepackage[T1]{fontenc}    % use 8-bit T1 fonts
\usepackage{hyperref}       % hyperlinks
\usepackage{url}            % simple URL typesetting
\usepackage{booktabs}       % professional-quality tables
\usepackage{amsfonts}       % blackboard math symbols
\usepackage{nicefrac}       % compact symbols for 1/2, etc.
\usepackage{microtype}      % microtypography
\usepackage{lipsum}
\usepackage{graphicx}
\graphicspath{ {./images/} }
\usepackage{multirow}
\usepackage{bbding}
\usepackage{pifont}
\usepackage{caption}
\usepackage{amsmath}
\usepackage{array}

\title{Data Cleaning of Data Streams\thanks{Preprint of the Springer Handbook of Data Engineering}}

\newbox{\orcid}\sbox{\orcid}{\includegraphics[scale=0.06]{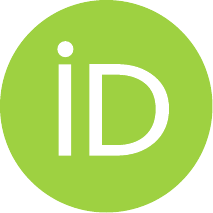}}
\author{
  \href{https://orcid.org/0000-0002-5960-5886}{\usebox{\orcid}\hspace{1mm}Valerie Restat} \\
  University of Hagen \\
  Hagen, Germany\\
  \texttt{valerie.restat@fernuni-hagen.de} \\
  \AND
  \href{https://orcid.org/0009-0002-3371-212X}
   {\usebox{\orcid}\hspace{1mm}Niklas Rodenhausen} \\
  University of Hagen \\
  Hagen, Germany\\
  \texttt{n.rodenhausen@googlemail.com} \\
  \And
  \href{https://orcid.org/0009-0009-3994-8038}
   {\usebox{\orcid}\hspace{1mm}Carina Antonin} \\
  University of Hagen \\
  Hagen, Germany\\
  \texttt{carina.antonin@gmail.com} \\
  \And
  \href{https://orcid.org/0000-0003-2771-142X}
   {\usebox{\orcid}\hspace{1mm}Uta St{\"o}rl} \\
  University of Hagen \\
  Hagen, Germany\\
  \texttt{uta.stoerl@fernuni-hagen.de} \\
}

\begin{document}
\maketitle
\begin{abstract}
Streaming data can arise from a variety of contexts. Important use cases are continuous sensor measurements such as temperature, light or radiation values. In the process, streaming data may also contain data errors that should be cleaned before further use. Many studies from science and practice focus on data cleaning in a static context. However, in terms of data cleaning, streaming data has particularities that distinguish it from static data. In this paper, we have therefore undertaken an intensive exploration of data cleaning of data streams. We provide a detailed analysis of the applicability of data cleaning to data streams. Our theoretical considerations are evaluated in comprehensive experiments. Using a prototype framework, we show that cleaning is not consistent when working with data streams. An additional contribution is the investigation of requirements for streaming technologies in context of data cleaning.
\end{abstract}

% keywords can be removed

\section{Introduction}
\label{sec:introduction}
Data is rarely error-free and data cleaning plays an essential role in ensuring data quality~\cite{Abedjan2016}. Insufficient data cleaning can lead to incorrect analysis results and unreliable decisions~\cite{Chu2016}. For this reason, a number of methods and tools for detecting errors and cleaning data exist. However, many of these methods are focused on static data sets~\cite{Chu2016}. To meet the high timeliness requirements of big data users, \emph{data streams} are often used in modern big data analysis applications. In order to develop reliable data cleaning methods for this type of data sets, some special characteristics of data streams need to be taken into consideration. 

Thus, in this paper, we conduct a comprehensive exploration of \emph{data cleaning of data streams}. Our theoretical considerations are practically evaluated in extensive experiments. The effects of the peculiarities of streams on different types of errors are analyzed. Subsequently, the requirements for technologies are examined.

The remainder of this paper is structured as follows. In Section~\ref{sec:data_streams} the special features of data streams that need to be taken into account when cleaning this type of data are discussed. Related work is presented in Section~\ref{sec:related_work}. Section~\ref{sec:dc_streams} addresses the applicability of data cleaning to data streams. The results are practically evaluated in Section~\ref{sec:evaluation} using a prototype framework. In addition, requirements for technologies are presented in Section~\ref{sec:technologies}. The contributions of this paper are then summarized in Section~\ref{sec:conclusion} and possible future work is discussed.

\section{Data Streams}
\label{sec:data_streams}
In order to highlight the specifics of data cleaning of data streams, in this section, the definition of a data stream is discussed first. In this work, a data stream is understood as an \emph{unbounded, ordered series of data vectors}. Each \emph{data vector} is provided with a \emph{timestamp}. This timestamp indicates the time when the data vector was transmitted to the data cleaning pipeline.

Following this description and in line with Kenda and Mladenić~\cite{Kenda2018}, a data stream $\mathbb{S}$ is mathematically defined as follows: 
 \begin{equation}
\mathbb{S} = \left \{ 
\left (
v_1=
\begin{bmatrix}
	x_{1,1}\\ 
	x_{1,2}\\ 
	...\\
	x_{1,q}
\end{bmatrix}
,t_1
\right )
,
\left (
v_2=
\begin{bmatrix}
	x_{2,1}\\ 
	x_{2,2}\\ 
	...\\
	x_{2,q}
\end{bmatrix}
,t_2
\right )
,...
,
\left (
v_n=
\begin{bmatrix}
	x_{n,1}\\ 
	x_{n,2}\\ 
	...\\
	x_{n,q}
\end{bmatrix}
,t_n
\right )
\right \}
\end{equation}
Where $1 < m < n$ and $m, n \in \mathbb{N} $. Further, $t_m$ is the time at which the vector $v_m$ arrives in the system. The \emph{dimension} of each vector $v_m$ is $(1, q)$ and $x_{m,p}$ is the \emph{measured value} for the attribute $p$ in the vector $v_m$. $t_n$ would theoretically be the time at which the last vector of the stream, $v_n$, arrives in the system. Due to the \emph{unboundedness} of a stream by definition, however, $t_n$ is never reached. A streaming data set is the collection of all previously known data vectors and their attribute expressions at a time $t_m$, formally $\mathbb{S}_m$.

Before arrival of vector $v_1$ in the stream, there is no information about the streaming data set beyond the data scheme of the vectors. For all vectors from $v_2$ to $v_m$, at least one but never all vectors of the stream are known in the streaming data set.

This entails the following particularities for data cleaning of data streams:
\begin{itemize}
    \item Upon arrival of vector $v_1$, no information about the streaming data set $\mathbb{S}_m$ beyond the data schema is available for data cleaning
    \item When cleaning all vectors except $v_1$, the streaming data set $\mathbb{S}_m$ consists of at least one vector
    \item Never all vectors of the data stream $\mathbb{S}$ are known. 
\end{itemize}

According to the proposed definition, a data vector has a \emph{fixed schema}. This is similar to a table in a relational database. As in a table, values can be unknown as long as no further restrictions are set. Thus $x_{k,p}$ can be NULL.

Different vectors can contain the information provided in the schema for different entities -- like sensors or vehicles -- respectively. This information arrives sequentially in the stream. A natural order of vectors can be derived from the time of arrival of a vector. Data streams therefore always have a \emph{time reference}. However, a stream is not necessarily a \emph{time series}. In a time series, the same entity is described at different points in time~\cite{Zhang2024}. A data stream, on the other hand, can contain information about different entities, as described above. Thus, a data stream is not equivalent to a time series. Rather, a time series is a special case of a data stream.

The formal definition of data streams presented would in general also include those data pipelines, where an incremental expansion of the data set occurs at larger time intervals (e.g., daily). In these scenarios, manual-iterative data cleaning by humans is conceivable. To distinguish this batch-based type of data loading process from streaming processes, this work assumes that a data stream must process data with the lowest possible \emph{latency}. Data vectors must be made immediately usable for analysis~\cite{Gomes2019}. Thus, manual-iterative data cleaning is not possible and there is a need for automated processing.

The need to produce near real-time insights with low latency sometimes results in further limitations. Stream applications often process a high volume of data~\cite{Roger2019}. To cope with data volumes and latency requirements, the most recent data history is often stored in memory. Thus, no access to disk resident data is necessary. This is often managed by \emph{windows}. Different approaches for windowing functionality exist~\cite{Gedik2014}. When using windows, the available information about the data set is further limited. Distribution-based characteristics are only obtainable within one window.

In summary, the following particularities result for data cleaning of data streams:
\begin{itemize}
    \item The streaming data set $\mathbb{S}$ and its statistical characteristics $\mathbb{C}$ are never fully known. Only the characteristics about the data set $\mathbb{S}_m$ are available at time $t_m$. When using windows, only the characteristics about the current window are available.
    \item With each new vector, the characteristics $\mathbb{C}$ about the data set or the window (e.g. mean or standard deviation) may change: $\mathbb{C}(\mathbb{S}_{m-1}) \neq \mathbb{C}(\mathbb{S}_m) \neq \mathbb{C}(\mathbb{S}_n)$. 
    \item To be able to analyze new data vectors immediately after arrival, data cleaning of vector $v_m$ must always take place at time $t_m$.
    \item Subsequent modification of vector $v_m$ is not possible, because the vector is already used for further analysis.
\end{itemize}

Derived from this, two properties can be determined that distinguish data cleaning for data streams from data cleaning for static data sets:
\begin{itemize}
    \item \textit{Time dependency}: Since the streaming data set is never completely known, the properties may change with each new vector. Important statistical metrics such as mean and standard deviation as well as the uniqueness of attribute values are thus not constant. They depend on the point of time of processing.
    \item \textit{Automated processing}: The vectors of a data stream must be cleaned immediately and made available for further applications. Manual-iterative data cleaning by humans is therefore hardly possible. The cleaning process must be fully automatable. In addition, once cleaning decisions have been comitted, they can no longer be corrected retrospectively, as the data vectors affected by these decisions have already been used in other subsequent applications..
\end{itemize}

This is clearly distinguished from static data cleaning. Here, the definitive statistical information about the data set is available. In addition, a higher level of human interaction is possible and often common. Hence, not all methods of static data cleaning are easily transferable to streaming data.

This section introduced the definition and particularities of a data stream. These are necessary in order to analyze the applicability of data cleaning. In the next section, the related work in the area of data cleaning of data streams is presented.

\section{Related Work}
\label{sec:related_work}
Data cleaning of static data sets has been studied in a variety of works (e.g. \cite{Ilyas2019, Abedjan2016, Chu2016, Krishnan2016}). In contrast, only few works can be found in the context of data cleaning of data streams.

Song et al.~\cite{Song2015} present an approach for cleaning of streaming data, based on the constraints on the maximum change in an observed value between two points in time. However, this approach is only applicable to data streams with time series properties.

In Volkovs et al.~\cite{Volkovs2014}, a framework for continuous data cleaning is presented. For this purpose, constraint rules for the identification and correction of errors are investigated.

Räth et al.~\cite{Rath2023} present a system for interactive data cleaning in real-time streaming applications. Some cleaning operators were implemented as a proof of concept. However, the system is only described briefly. A detailed and comprehensive analysis of the different error types and cleaning methods is not available.

Mirzaie et al.~\cite{Mirzaie2023} have conducted a systematic literature review on the quality control of data streams. However, the focus here is more on the evaluation of data quality and not on data cleaning of data streams.

Various methods for cleaning streaming data are also discussed in the context of Wireless Sensor Networks (WSN)~\cite{Sandric2019}. WSNs are networks that are located in close proximity to each other. Data cleaning is based on the assumption that measurements from sensors that are close to each other typically do not differ much.

In the context of mobile WSNs, Tasnim et al.~\cite{Tasnim2017} also present a cleaning approach. They propose a method based on weighted moving average to better determine missing values.

However, all of this work focuses on specific methods or application areas. A fundamental investigation of the data cleaning of data streams could not be found.

\section{Data Cleaning and Applicability to Data Streams}
\label{sec:dc_streams}
This section takes a closer look at data cleaning and its applicability to data streams. In order to do this, the general data cleaning process and its methods are first described. Subsequently, different error types are presented and their applicability to data streams is analyzed. The particularities of data cleaning of data streams are derived from this and summarized at the end of this section. 

Data cleaning refers to the detection and repair of errors in data~\cite{Ilyas2019}. An error exists in the data if the measured or observed value in the data set deviates from the true value. However, the true value of an attribute expression is not known, because otherwise data cleaning would not be necessary. For this reason, criteria or rules are defined in a data cleaning process according to which a value is defined as erroneous. The goal of data cleaning is therefore primarily to bring a data set into a state that satisfies all decision rules relevant to error detection. While repairing, the corrected value should be as close to the true value as possible. However, since the true value is not known, the aim of the repairing is to achieve as good an approximation as possible. Outside of experimental contexts where ground truth is known, this approximation is not verifiable. Therefore, the algorithms for repairing are primarily based on theoretical considerations and assumptions.

The following describes in more detail the data cleaning process, associated methods, and possible data errors. In particular, the special features that arise in data cleaning of data streams are discussed.

\subsection{Data Cleaning Process}
\begin{figure}[t]
\centering
% \sidecaption[t]
\includegraphics[scale=.45]{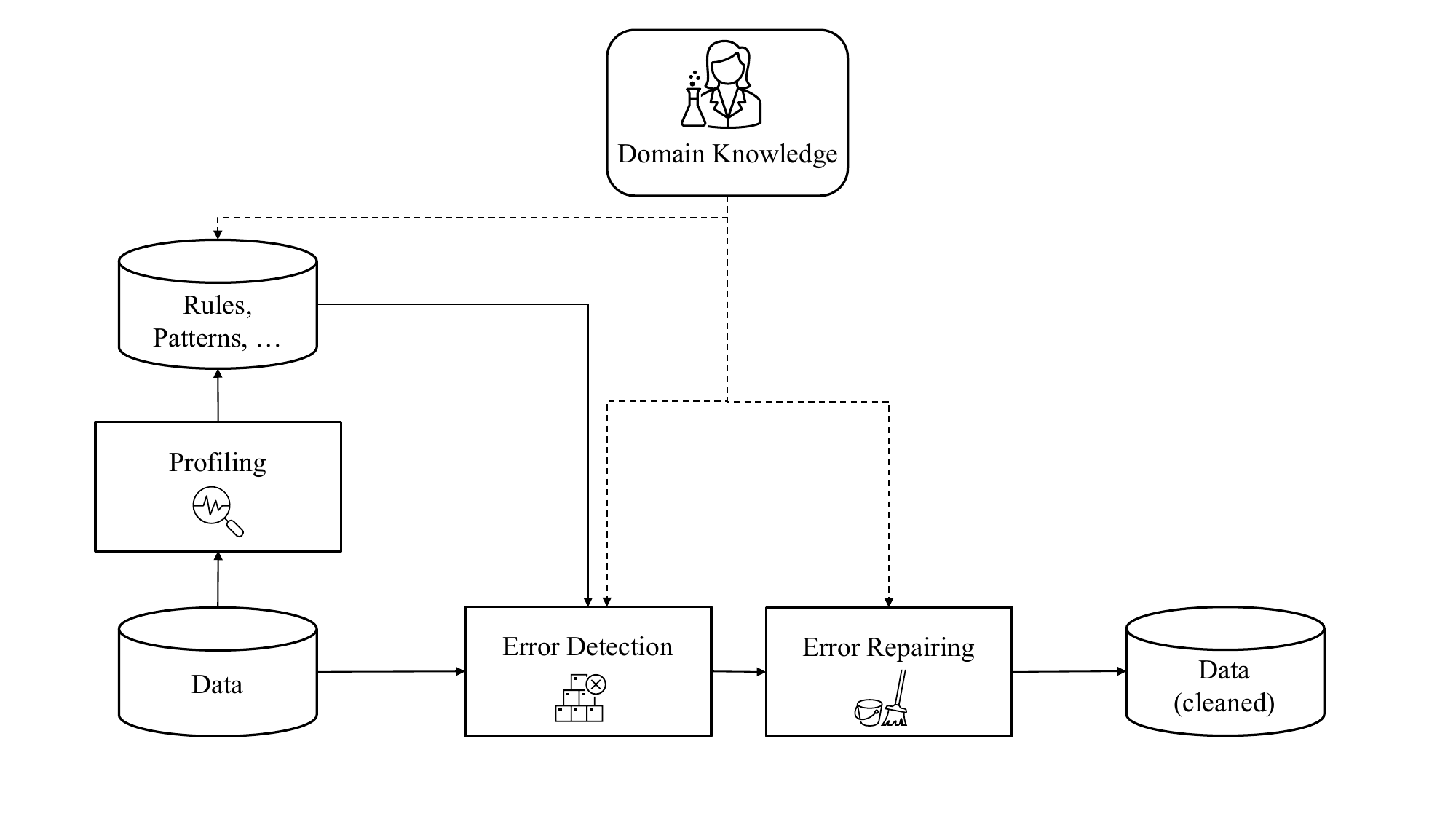}
\caption{Data Cleaning Process, based on~\cite{Ilyas2019}}
\label{fig:data_cleaning_process}
\end{figure}

Figure~\ref{fig:data_cleaning_process} describes a data cleaning process, based on Ilyas and Chu~\cite{Ilyas2019}.
In a first step, data profiling is performed. Rules and patterns are derived from the data and statistical information is determined. Based on this, errors are detected in the data. In the error repairing step, these errors are cleaned. This process can be repeated iteratively.

\subsection{Data Cleaning Methods}
In the literature different approaches for the classification of data cleaning methods are described. 

One possibility is rule-based methods. A distinction can be made as to whether such rules are checked interactively or automatically. Interactive data cleaning methods require human involvement in the data cleaning process. Different groups can be be included. This involves domain experts who are very knowledgeable about the application domain of the data, but also knowledge bases and crowdsourcing~\cite{Chu2015}. Automated data cleaning methods can perform the data cleaning process without human interaction. The rules are either derived in advance by domain experts and stored in a formalized way or are also generated automatically by an analysis of the data set. The scientific literature also explores data cleaning approaches in which the automated decisions of the system are reviewed by humans at intervals and the results of the review are in turn fed back into the automated cleaning workflow~\cite{Mahdavi2019}. Since a property of data streams is automated processing, only automated data cleaning methods can be used accordingly.

Model-based methods are another option. These use machine learning techniques to detect and repair errors in the data. One possible approach is to develop and train a model of the data creation process. For incorrect data sets, the model generates correction values based on probabilistic considerations and the correlations in the data set depicted in the model. With such model-based approaches, solutions must be found for the time dependency of data cleaning in data streams. A model of the data generation process can only contain the structures and correlations that are in the data set at the time of training. Moreover, model-based approaches usually require extensive, annotated training data. The lack of availability of a sufficiently large set of training data, as is usually the case with real data cleaning applications, is a major challenge when developing a productive data cleaning process based on model-based techniques~\cite{Ilyas2019}.

\subsection{Data Errors}
\label{sec:data_errors}
Data can contain a range of different errors~\cite{gouda-paper}. In this work, two limitations come into play when describing possible errors: In Section~\ref{sec:data_streams}, streaming data was defined as a sequence of data vectors where each vector has the same attributes. For this reason, only structured data is taken into account. Error types that can only occur with semi-structured or unstructured data are not considered in this work. This is an aspect for future work. In addition, only one data stream is examined as a source in the first step. Accordingly, errors that only occur when merging multiple data sources are not considered.

Taking these two limitations into account, nine types of errors were selected from the set of all possible errors for closer examination. Table~\ref{tab:error_classification} gives an overview of the selected error types. They are described in more detail below.

\begin{table}
\centering
  \caption{Error classification, based on~\cite{Rahm2000}}
  \label{tab:error_classification}
  \begin{tabular}{ll}
    \toprule
    \multirow{4}*{\emph{Schema Level}} & Uniqueness violation\\
     & Wrong data type\\
     & Interval violation\\
     & Functional dependency violation\\
     \midrule
     \multirow{5}*{\emph{Instance Level}} & Missing Values \\
      & Terminology heterogeneity\\
      & Duplicates\\
      & Outlier\\
      & Contradicting records\\
  \bottomrule
  \end{tabular}
\end{table}

The errors are divided into the two groups \emph{Schema Level} and \emph{Instance Level}, based on the classification of Rahm and Do~\cite{Rahm2000}. A schema contains freely defined rules for the data set. An error at the schema level would not lead to content inconsistencies in the data set. Rather, for errors of this class, data vectors contradict formal expectations for the data. A schema-level error occurs when an attribute expression (or combination of attribute expressions) contradicts the rules defined in the schema~\cite{Rahm2000}. Automated detection and repairing therefore appears feasible. However, the decision as to whether the cause of an error at schema level is an incorrect attribute specification or an inadequate schema is usually only possible on the basis of in-depth investigations. Thus, this can only be automated to a limited extent. For an automated repairing of an error at the schema level, it is therefore necessary to determine whether the repairing is to be made by changing the attribute value or by changing the schema.

Errors at the instance level lead to inconsistencies within a data set. They thus compromise the usability of the data. For example, information could be missing in a vector which is always present in the observation context. Furthermore, redundancies could arise in the data set or data points could differ conspicuously from the rest of the data set and thus the application domain. Due to the content reference, it is to be expected that an automated identification, but especially a proper correction of instance-level errors is difficult to implement in the streaming context.

\subsubsection{Uniqueness violation}
Uniqueness is a requirement defined in the schema that each value of an attribute may only occur exactly once in the data set. This concerns attributes that uniquely identify an entity. Examples are a serial number or any comparable, generated ID value. The requirement can also refer to a combination of several attributes.

\begin{description}
    \item[\textit{Detection}] To detect a uniqueness violation, a check is made for each value of the attributes in question to see if it exists only once in the data set.
    \item[\textit{Repairing}] Uniqueness constraints are often used for attributes that uniquely identify entities or facts. Therefore, in case of a uniqueness violation, an arbitrary value not previously present in the data set cannot be used as a substitute. Even if this correction complied with the formal requirement of uniqueness constraints, it would lead to a contextual error. Vectors that violate the uniqueness requirement are therefore usually rejected with an error message when they are included in the data set~\cite{Rahm2000}. In static data sets, a repair can be made by removing the record(s) that violate the uniqueness requirement. A rule must be created to decide which record to remove. This usually requires domain knowledge.
    \item[\textit{Applicability to data streams}] In a data stream, new data vectors are successively added to the data set. In order to detect uniqueness violations, when a new data vector is added, it must be checked whether the value of the attributes for which a uniqueness requirement applies already exists in the data set. Due to the time dependency, this can lead to different results in the cleaning depending on the order of the incoming vectors. The vector arriving first for processing is necessarily always considered error-free. 
\end{description}

\subsubsection{Wrong data type}
In many software-based applications, specified data properties are expected during application development. These properties can be defined by specifying a valid data type in the schema. The data type of an attribute defines, which types of operations are supported. If the expected properties are not met, there is a high risk of errors in the further data processing pipeline due to invalid operations.

\begin{description}
    \item[\textit{Detection}] Unless the data generation process can guarantee compliance with the expected data type, verification should be done by analyzing the values. This analysis can be performed with the help of user-defined program code or by using type conversion functions available in many programming languages. In some cases, the data type may also be included in the data transfer format as additional syntactical information.
    \item[\textit{Repairing}] Attribute values with wrong data types can be converted to the required data type using the type conversion functions. If conversion using this functions is not possible by these methods, typical problems that hinder conversion can be fixed in a second step. These problems include using a comma instead of a period for floating point values, or using a different regional syntax for date-related values. If conversion is still not possible, the incorrect value has to be removed and, if necessary, a replacement value with the correct data type must be generated.
    \item[\textit{Applicability to data streams}] For automatic generation of data in streaming pipelines, the data types are often already defined in the data generation software. Therefore, it can be assumed that a measured value with a wrong data type already triggers an error on the part of the data generation process. Nevertheless, it is still useful to check all incoming vectors for schema-compliant data types. In this way, transmission errors can be detected and manual entries can be checked to ensure further processing in the streaming pipeline.
    
    This type of error refers only to individual attribute values. No knowledge about other values in the data set is required. Detection and  repairing can be automated easily. Time dependency and automation requirements do not affect this type of error.
\end{description}

\subsubsection{Interval violation}
Intervals can be discrete or continuous. For a discrete interval, the value of an attribute must be part of a list of allowed values. For continuous attributes, the value of an attribute must be within two boundaries that define the value range. The definition of permissible intervals requires knowledge about the facts represented in the data, the conditions of its generation, and the measurement procedure.

\begin{description}
    \item[\textit{Detection}] In order to detect interval violations, it must be checked wether the attribute values are within the permissible range. The allowed values for an attribute are defined in the schema of the data set either in the form of a list or by limit values. The check can be done algorithmically and is unproblematic even with complex interval definitions. By defining the interval in the form of formal rules, an automated detection of deviations can therefore be realized~\cite{Ilyas2019}.
    \item[\textit{Repairing}] Repairing can be performed by correcting the observed value or by extending the interval. Assuming that the observed value is an approximation of the true value, at least for ordinally scaled attributes, using the closest value that lies within the permissible interval is a possible strategy for determining a substitute value. The situation is different if the observed value cannot be regarded as an indicator of the true value. In this case, distribution measures such as the mean, but also values from model- or distance-based methods can be applied. In addition, it is also possible to generate random substitute values within the valid interval.
    \item[\textit{Applicability to data streams}] Interval violations can occur in the streaming context due to various causes. Examples are measurement and transmission errors or structural changes of the data generating process over time. Detection can be performed automatically without knowledge of the values of other vectors.

    For repairing, values based on distributions of the data set can be determined in the streaming context, taking into account the time dependence of these values. The distribution can only be determined based on the available data set $\mathbb{S}_m$. For this purpose, either the complete data set $\mathbb{S}_m$ or a window over a defined range of previous values can be used. However, the distribution can change with each new vector. The value for repairing thus depends on the position in the data stream. Repair using random values within the valid interval is possible without restrictions.
\end{description}

\subsubsection{Functional dependency violation}
Functional dependencies (FDs) are restrictions defined in the data schema according to which permissible values for an attribute in a vector functionally depend on the value of one or more other attributes of this vector~\cite{Papenbrock2016}.

\begin{description}
    \item[\textit{Detection}] Functional dependencies can be defined during the development of the data streaming pipeline from domain knowledge or by rules in a data set~\cite{Ilyas2019}. Compliance with functional dependencies can therefore be checked with the help of a formal set of rules~\cite{Ilyas2015}. However, it is also possible that the functional dependencies are not available as rules. In this case, discovery algorithms are needed to uncover them in a given data set~\cite{Papenbrock2016}.
    \item[\textit{Repairing}] Repairing functional dependencies can be done by adjusting data, redefining rules, or adjusting data and rules in parallel.
    \item[\textit{Applicability to data streams}] Functional dependencies can be detected and repaired automatically. If the functional dependencies are known, detection of violations can be performed at the time of arrival of a vector without knowledge of other data vectors. Therefore, the time dependency does not affect the detection. Structural changes in the data generation process can result in correctly transmitted data vectors, which, however, contradict existing functional dependencies. In data streaming pipelines, it must therefore be possible to adjust functional dependencies. A manual adjustment as part of a maintenance routine, where the effects of the change on subsequent applications can be analyzed, appears preferable to an automated adjustment process. However, if the functional dependencies are not available as rules, the detection is significantly affected by the time dependency. Discovery algorithms normally require the entire data set in order to find valid functional dependencies. In the streaming context, though, the entire data set is never available.
\end{description}

\subsubsection{Missing Values}
This error type can occur for various reasons, including incorrect manual entries, errors during data acquisition or sensor failures~\cite{Kaiser2014}. Another form of this error type is completely missing data vectors. Missing values do not necessarily represent an error that requires repairing. Sometimes they are defined as permissible by the generation process. Nevertheless, they complicate further processing of the data. Some models and statistical methods can only deal with complete data or they have only simple strategies for dealing with missing values. This includes deleting incomplete data vectors, which can result in information loss.
\begin{description}
    \item[\textit{Detection}] In the simplest case, the value of the attribute is \emph{NULL} and can therefore be easily detected. More difficult are missing values that are encoded by specific values. These can be placeholders like \emph{N/A} or \emph{-9999}. It may also happen that a default value is inserted instead of \emph{NULL}. In general, the classification of certain values as markers for missing values can be mapped in an automated set of rules~\cite{Barateiro2005}. But these markers have to be stored explicitly. Since a large number of placeholders are conceivable, it is hardly possible to map all possible variants. This task requires knowledge of the domain and the data generation process. Sufficient knowledge about the data generation process is also needed to detect missing data vectors (e.g. the process generates one vector generated every 5 minutes). To identify missing data vectors, a comparison of the existing vectors with the expected vectors must be performed using an appropriate index. A logical rule can usually be used for this purpose, allowing automated identification.
    \item[\textit{Repairing}] Different classes of methods exist for repairing missing values~\cite{Kaiser2014}. One option is to remove the missing values. However, this could easily lead to loss of information. Another approach, which does not require the calculation of replacement values, is to process missing values using a separate ruleset~\cite{Kaiser2014}. Repair can also be done by replacing with distribution-based values. Typical replacement values are mean, median or mode. For such values, information about the entire data set is necessary. In addition, repairing is also possible with techniques based on statistical methods or machine learning. Depending on the context, information from other vectors or even the whole data set is needed. For data sets with a natural order, such as time series, interpolation methods can be applied for error repairing~\cite{Bruijn2016}.
    \item[\textit{Applicability to data streams}] Missing values can occur in both static and streaming data sets. If an entry is created automatically, which can usually be assumed for streams, then missing values due to manual entry are unlikely to appear. However, missing values can occur due to scenarios such as the failure of measurement sensors. The detection of missing values can be easily automated in data streams based on a list of feature occurrences classified as erroneous. Knowledge of other values in the data set is not required for this purpose. Therefore, time dependence is not relevant in identifying missing values. The detection of undefined placeholders as missing values, on the other hand, is difficult to automate. Repairing missing values by correcting them without replacing the values is easy in streaming applications, provided the target application can compensate for the loss of information. Distribution-based values can only be determined based on the already known streaming data set $\mathbb{S}_m$ due to the time dependence. This is similar to repairing interval violations using distribution-based methods. Similarity-based methods are applicable in the streaming context when a suitable comparison data set is available. Latency requirements may not be met for compute-intensive processes. For model-based approaches, a sufficient training data set must be available as described. In the special case of time series, interpolating approaches can be used. However, it is a prerequisite for usage of these methods in streams that only the previous and not the not yet available subsequent vector is referred to.
\end{description}

\subsubsection{Terminology heterogeneity}
Terminology heterogeneity is present in the data when the same real-world observation is represented by different values. Errors of this type occur especially with non-numeric data types. Based on Oni et al.~\cite{Oni2019}, in this work the following error types are grouped under terminological heterogeneity:
\begin{itemize}
    \item \textit{Spelling mistakes}: The representation of a term contains a syntactic error, e.g. ``cleaning'' is written as ``cleening''.
    \item \textit{Heterogeneous terms}: A term is mapped to different, syntactically correct representations.
    \item \textit{Divergent scales}:. An observation is represented by differently scaled values.
    \item \textit{Different grouping}: An observation is classified into different groups.
    \item \textit{Heterogeneous data}: A value is represented in different formats, e.g. 03/22/2019 vs. 20190322.
\end{itemize}

In highly automated data generation processes, which are often the basis of streaming pipelines, terminological heterogeneity in a single stream can be considered unlikely.

\begin{description}
    \item[\textit{Detection}] Detection depends on the specific type of terminology heterogeneity. They all have in common that they require a high degree of human interaction. For example, if the scale is not specified for type \emph{divergent scales}, no error can be detected in the data. Methods such as outlier detection can only provide indications for the detection of deviating scales. A final assessment would have to be carried out by a domain expert.
    \item[\textit{Repairing}] Repairing also depends on the type of heterogeneity error. This usually requires extensive rules, e.g. to standardize different groupings or to determine a uniform syntax for heterogeneous data.
    \item[\textit{Applicability to data streams}] The detection and repair of the different types of terminological heterogeneity using automated algorithms seems difficult or impossible to implement. The interpretation of the semantic content of information, which is usually necessary, can hardly be realized without human intervention. However, manual iterative data cleaning by humans is not feasible in automated streaming pipelines. Therefore, terminological heterogeneity should be excluded as far as possible in the data generation process, e.g. by limiting the possible attribute values in a data entry process. For the data streaming pipelines studied in this work, the error type of terminological heterogeneity is of minor importance due to the mostly automated data generation by sensors or computer programs. Therefore, the cleaning of terminological heterogeneity in data streaming pipelines is not investigated further in this work.
\end{description}

\subsubsection{Duplicates}
A duplicate in a narrower definition exists if several data vectors contain the same attribute characteristics. In a broader definition, a duplicate also exists if the same information is available in the data set for an observation entity, but it is linked to different key attributes~\cite{Rahm2000}.

\begin{description}
    \item[\textit{Detection}] Detection of duplicates in a narrow definition can be done by comparing all values of a data vector with the values of all other vectors in the data set. If all attribute values in several data vectors are identical, than these vectors must be seen as duplicates. The detection of duplicates in a broader sense, is considerably more difficult. In this case, possible duplicates can be found by comparing the values of all attributes in a data row without considering the primary keys. However, the decision whether a duplicate exists in the case of identical non-key attributes or whether it is simply a matter of two observations with the same properties depends on the application domain. In general, a more in-depth human inspection is required.
    \item[\textit{Repairing}] Duplicates in the narrower sense are repaired when there are no longer two data vectors with identical attribute expressions. Therefore, a simple way to repair is to remove all redundant data vectors and keep only one arbitrary vector. In the case of duplicates in the broader sense, the options for repair are identical to the case of duplicates in the narrower sense, once the primary keys have been aligned. The particular challenge lies in the identification of this type of duplicates. It can be assumed that the investigations of the data vectors associated with this detection require manual iterative cleaning steps.
    \item[\textit{Applicability to data streams}] In the streaming context, detection and repairing must be automatic, taking into account the time dependency. In this context, the issue already discussed for error type \emph{uniqueness violation} also occurs with duplicates. The vector that arrived first in the data cleaning process must automatically be considered the correct version of the duplicate vectors.
\end{description}

\subsubsection{Outlier} According to Hawkins~\cite{Hawkins1980}, an outlier is an observed value that deviates from other observed values to such an extent that this value is assumed to have occurred by a different mechanism than all other values. One difficulty in detecting outliers is defining ``normal'' values and deciding at what distance a value is no longer considered ``normal'' but an outlier~\cite{Ilyas2019}. However, cleaning outliers is not always expedient. Outlier detection can also be the primary goal of data analysis, such as fraud detection or monitoring of technical equipment.

\begin{description}
    \item[\textit{Detection}] Outlier detection is a much discussed topic in the literature. Ilyas et al.~\cite{Ilyas2019} distinguish three groups of methods: statistical methods, distance-based methods and model-based methods. In addition, different methods are presented in the literature for time series that exploit the time component for outlier identification~\cite{Thakkar2016}. Since this topic is so extensive, we do not discuss each method in detail here. Instead, reference is made to literature such as~\cite{Chandola2012} and ~\cite{Thakkar2016}.
    \item[\textit{Repairing}] A replacement value can be generated in different ways. The data point can be replaced by a distribution value of the data set, such as mean or median. Such central distribution values are usually not classified as outliers. Alternatively, a replacement value can be defined using the calculation logic of the identification methods. In this case, a substitute value is used that comes as close as possible to the measured, conspicuous value, but is no longer classified as an outlier by the selected identification method. Outliers can also be removed from the data set. However, this may affect important properties of the data set, such as distributions, for later analysis.
\end{description}

\subsubsection{Contradicting records}
Contradictory data exist when there are different values for an attribute that cannot be valid at the same time~\cite{Rahm2000}. The error type is very similar to the error types \emph{duplicates} and \emph{uniqueness violation}. Therefore, similar detection and repairing mechanisms can be used. The main difference is that \emph{contradicting records} as instance-level errors already causes logical inconsistencies in the data set without explicit rule definition. \emph{Duplicates} and \emph{uniqueness violation} are based as errors on schema level on freely defined rules without reference to content~\cite{Rahm2000}.

\begin{description}
    \item[\textit{Detection}] Automatic detection is easily possible if both the identifying attributes and the attributes that may have only a single value for the specific entity are specified at the same time when defining the data schema.
    \item[\textit{Repairing}] To resolve the contradiction, an entity with a key attribute may only have different values for those attributes for which this is permitted by the requirements of the domain. Alternatively, only one of the data vectors with this key attribute can be selected and all other affected vectors removed. In many data cleaning tools, the repair of conflicting data must be performed by a human user~\cite{Barateiro2005}.
    \item[\textit{Applicability to data streams}] The rules for detecting contradicting records can usually be represented algorithmically. For this purpose, each new vector is compared with already existing vectors on the basis of its key attributes and removed if contradictions exist. The automation of the repair can therefore be assumed for this type of error. However, a peculiarity arises from the time dependence. The first vector of a group of contradictory vectors must always be classified as correct, even if it turns out due to later available information that already this first vector is incorrect. The only repair options are to align new vectors with their counterpart already present in the data set or to remove the new vectors.
\end{description}

\subsection{Particularities of Data Streams}
In the following, the effects of peculiarities of data streams are summarized depending on the error types presented.

\subsubsection{Effects of time dependency}
As described in Section~\ref{sec:data_streams}, the characteristics $\mathbb{C}$ of a streaming data set $\mathbb{S}$ are never completely known and can change with each new data vector. However, many of the data cleaning approaches available in the literature and in practice are based on distributional properties. In streaming applications, these properties are not constant. Table~\ref{tab:time-dependent} summarizes the effects of the time dependency on the error types.
\begin{table}
\centering
  \caption{Effects of time dependency -- \ding{55} indicates that the error type is affected by time dependency}
  \label{tab:time-dependent}
  \begin{tabular}{lcc}
    \toprule
    Error Type & \multicolumn{2}{c}{Affected by time dependency} \\
    \midrule
      & Detection & Repairing \\
     \emph{Schema Level} & & \\
     \hspace{3mm}Uniqueness violation & \ding{55} & \ding{55}\\
     \hspace{3mm}Wrong data type & - & -\\
     \hspace{3mm}Interval violation & - & -\\
     \hspace{3mm}Functional dependency violation & - & -\\
     \emph{Instance Level} & & \\
     \hspace{3mm}Missing values & - & \ding{55}\\
     \hspace{3mm}Terminology heterogeneity & n/a & n/a\\
     \hspace{3mm}Duplicates & \ding{55} & \ding{55}\\
     \hspace{3mm}Outlier & \ding{55} & \ding{55}\\
     \hspace{3mm}Contradicting records & \ding{55} & \ding{55}\\
  \bottomrule
\end{tabular}
\end{table}

For error types \emph{duplicates}, \emph{contradicting records}, and \emph{uniqueness violation}, the time dependency implies that data vectors with the corresponding attribute values that arrive first are always classified as correct. Each subsequent vector with colliding attribute values is always classified as incorrect. The decision as to which a vector is correct is thus made solely on the basis of the order of appearance in the data cleaning process. Contextual information or in-depth investigations, which are used to select the most suitable vector in manual-iterative static cleaning, cannot be used in streaming. The presence of an error is not noticed until the later vector appears in the data cleaning process. The earlier vector is then already processed and can therefore no longer be corrected.

For the repair of these error types, the time dependency has a different effect, depending on the method. The removal of incorrect values is independent of information from the data set. Thus, it is not affected by time dependency. However, removing it can cause information to be lost. Therefore, in most cases, a method should be chosen that finds a valid replacement value. If the algorithms used for replacement are based on the values of a data vector alone, without reference to the entire data set, they are not affected by time dependency. The situation is different if distribution-based algorithms are used. These are affected by the time dependency in any case. Characteristic distribution values change with each new vector. Furthermore, as described, the first vector is always necessarily correct. Thus, only subsequent vectors are considered for repairing.

With the error type \emph{outlier}, the classification of a data vector depends on the already existing vectors. This results in two effects. First, the classification of a vector as an outlier depends on the known state of the distribution measures of the streaming data set. A vector can therefore, depending on the state of the distribution measures at the time of cleaning, be classified once as an outlier and once not as an outlier. On the other hand, the corrections to earlier vectors, due to their influence on the distribution metrics of the data set, affect the classification of later vectors as outliers. As a consequence, insufficient or erroneous cleaning mechanisms influence the accuracy of the detection and repairing steps and, in the worst case, the data quality is worsened rather than improved by cleaning.

The detection of error types \emph{wrong data types}, \emph{interval violation}, \emph{functional dependency violation}, and \emph{missing values} is possible in most cases without reference to the streaming data set. The observation of the attribute values within the data vector is sufficient. However, distribution-based algorithms are often used especially for the repairing of missing values, e.g. mean and median imputation but also model-based approaches. These are affected by the time dependency as described.

It is shown that when developing a data cleaning pipeline for streaming data, detection and repair methods that require values of the data set should be avoided as much as possible. However, the use of time independent algorithms is not possible for all error types, especially not for the identification of the error types \emph{duplicates}, \emph{contradicting records} and \emph{uniqueness violation}. Proper cleaning of these error types in the streaming context is not possible because the first arriving vector must always necessarily be considered the correct vector. These three error types must therefore be avoided as much as possible during data generation and considered separately during data use in streaming context.

As discussed in Section~\ref{sec:data_errors}, instance-level errors in particular are affected by the peculiarities of the streaming context. These types of errors have a stronger reference to the entire data set. In contrast, errors at the schema level have no context in terms of content. For this reason, they are less affected by the specifics of data streams.

\subsubsection{Effects of automated processing}
Data cleaning in data streaming pipelines must be fully automated. Manual-iterative cleaning by humans is mostly not feasible. Table~\ref{tab:automated} summarizes the effects of automated processing on the error types.
\begin{table}
\centering
  \caption{Effects of automated processing -- \ding{55} indicates that the error type is affected by automated processing}
  \label{tab:automated}
  \begin{tabular}{lcc}
    \toprule
    Error Type & \multicolumn{2}{l}{Affected by automated processing} \\
    \midrule
      & Detection & Repairing \\
     \emph{Schema Level} & & \\
     \hspace{3mm}Uniqueness violation & - & -\\
     \hspace{3mm}Wrong data type & - & (\ding{55})\\
     \hspace{3mm}Interval violation & - & -\\
     \hspace{3mm}Functional dependency violation & - & -\\
     \emph{Instance Level} & & \\
     \hspace{3mm}Missing values & - & -\\
     \hspace{3mm}Terminology heterogeneity & \ding{55} & \ding{55}\\
     \hspace{3mm}Duplicates & - & -\\
     \hspace{3mm}Outlier & (\ding{55}) & (\ding{55})\\
     \hspace{3mm}Contradicting records & - & -\\
  \bottomrule
\end{tabular}
\end{table}

The need for automated processing precludes any data cleaning procedures that rely on human interaction and iterative error assessment. Rule-based approaches are therefore particularly suitable for data cleaning in a streaming context. Model-based approaches are suitable if a data set that is independent of the stream is available for building the models. In general, all methods for static data sets that can be algorithmically represented as decision and optimization problems can also be applied to streaming data sets. In automated form, such algorithms can even simplify and accelerate data cleaning of static data sets. The main difference is that in static data cleaning, trial-and-error~\cite{Mahdavi2019} is easily possible. When building data streaming pipelines, conceptual theoretical considerations have greater relevance.

With \emph{wrong data types}, the repairing can only be automated to a limited extent. Conversion rules are required. Also limited is the detection and repair of \emph{outliers}. Fixed boundary values are required for detection. The repair is restricted by an inflexible method selection. As described, automated processing is hardly feasible for detection and repair of \emph{terminology heterogeneity}. This error type is of minor importance due to the mostly automated data generation. It is thus not further examined in this work.

\bigskip
In this section, the data cleaning process and its methods were presented and the detection and repair of different types of errors was analyzed. The applicability to data streams was also discussed for each error type. It was shown that especially two particularities must be taken into account when working with data streams: the effects of time dependency and the effects of automated processing.

\section{Practical Evaluation}
\label{sec:evaluation}
The presented findings were practically evaluated in a prototypical framework. The design of the framework and the results of the experiments are presented in this section. Initially, the focus was on the impact on data quality. Therefore, the use of windows was dispensed with at this point. This allows for better investigation of the effects of time dependency in particular. By using windows, the effect of time dependency is further enhanced, as explained in Section~\ref{sec:data_streams}. In future work, we will investigate more deeply how the size of windows affects the results. In the first step, however, we have opted for a more general analysis. The necessity of stateful cleaning functions is presented in Section~\ref{sec:technologies} in connection with data streaming technologies. 

\begin{figure}[t]
\centering
% \sidecaption[t]
\includegraphics[scale=.35]{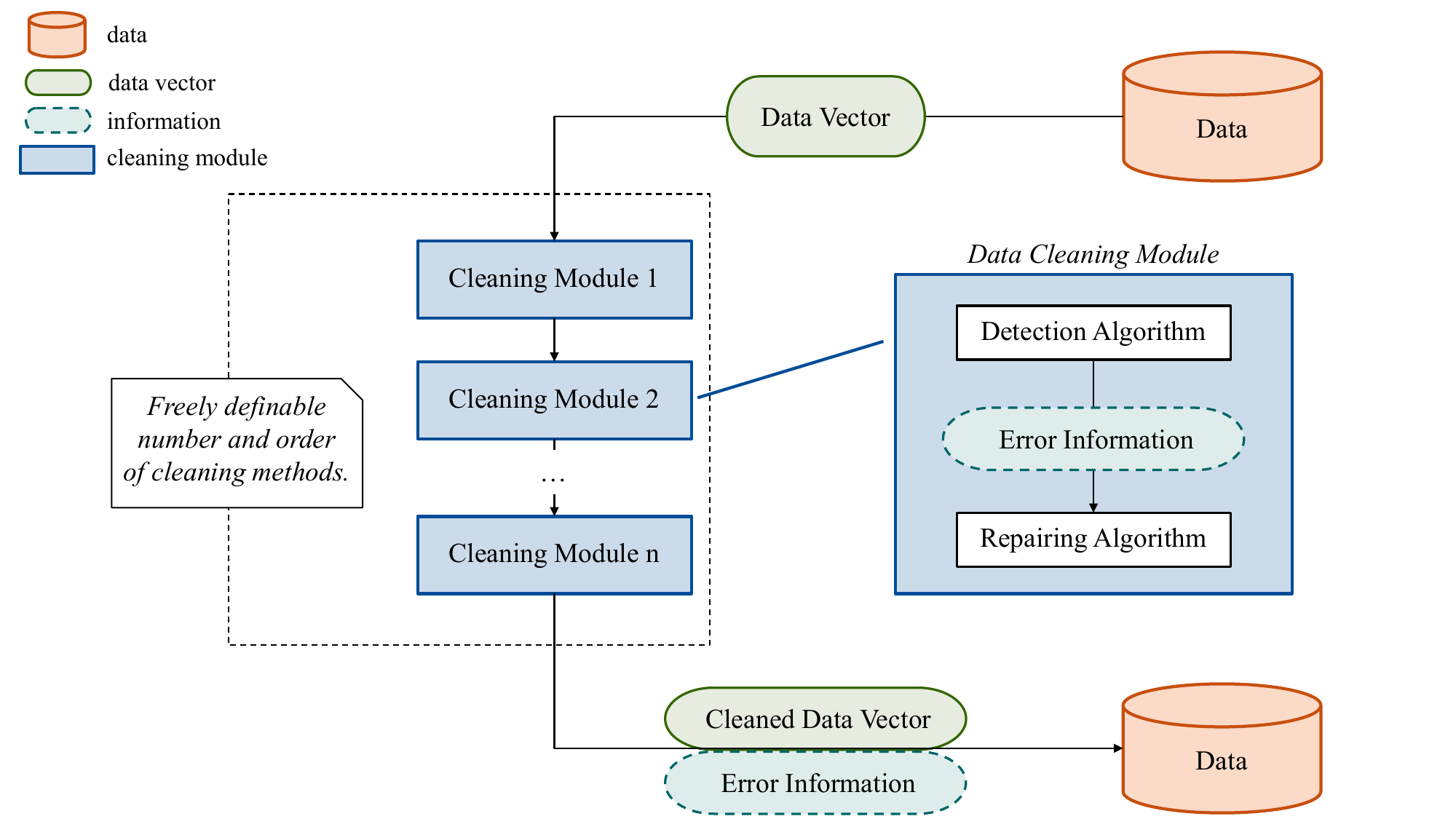}
\caption{Prototypical Framework for Data Cleaning of Data Stream}
\label{fig:framework}
\end{figure}

The presented results were practically evaluated in a prototypical framework. This is shown in Figure~\ref{fig:framework}.
It consists of a total of seven modules. With the exception of the excluded error  \emph{terminology heterogeneity}, a module has been developed for each error type presented in Section~\ref{sec:data_streams}. During testing of the prototype, it was determined that the error type \emph{wrong data types} cannot be used as an optional cleaning module due to the high importance of correct data types for automated processing. The algorithms of most cleaning modules require certain data types. Therefore, the checking of the data vectors for correct data types cannot be done in a cleaning module that is only optional. Instead, it takes place in a separate schema check module already before the first cleaning module.

The cleaning modules each consist of an detection step and a repairing step, which operate independently of each other and can be flexibly combined when building the pipeline. This allows the framework to be easily adapted to a specific use case and extended with additional algorithms. The repairing step can be omitted. In this case, only error detection is performed.

\subsection{Experiment Setup}
In this section, the data sets and the procedure of the experiment are described.

\subsubsection{Data Sets}
For the experiment, two test data sets were used, which contain data from real measurement and observation processes:
\begin{itemize}
    \item Intel Lab Data Set~\cite{IntelLab2015}
    \item New York Taxi Data Set~\cite{taxidataset}
\end{itemize}

The Intel Lab Data Set is a time series data set. The data was generated fully automatically by sensors over a period of six weeks. It contains measurement results for temperature, light intensity, humidity and voltage intensity. A version of the data set already preprocessed by Bruijn et al.~\cite{Bruijn2016} is used for the experiment. It contains data for 10 of the 54 sensors. This limitation allows for a more performant execution of multiple iterations of the experiment. The selected sample includes 17,695 measurement points within seven days. Thus, 87.7\% of the expected 20,160 measurement points are present in the data set. According to Bruijn et al.~\cite{Bruijn2016}, linear interpolation of the missing values is reasonable due to the short time periods with missing values. After interpolation, the data set contains the expected 20,160 measurement points in the selected time period.

The New York Taxi Data Set contains information about cab rides in New York City. As the individual cabs represent different entities, this is not a time series data set. The data is generated by interaction of the cab drivers with the cash register systems of the vehicles and is partially automated. It includes origins and destinations, fare, and various fare components. When preparing the data set, unnecessary columns were removed and differences in the summation of the cost items, which arose due to the use of two different cash register systems were eliminated. Only a sample is used for more perfomanent processing. This includes all entries with \emph{tpep\_pickup\_datetime} on march 1st 2022, which corresponds to 108,928 data vectors. This can be processed with the available resources with reasonable runtimes. The data set does not have a timestamp suitable for streaming for the time at which a data vector is loaded into the streaming pipeline. Therefore, such a timestamp is additionally generated. It is set to a random time between 5 and 60 seconds after \emph{tpep\_dropoff\_datetime}. For vectors with \emph{cancel\_flag = "cancel\_out"} an additional 60 seconds will be added to ensure that entries for canceled trips always arrive after the entry for the associated trip. The individual columns of the data set are explained in a data dictionary~\cite{taxidictionary}. Value ranges and constraints were derived from this data dictionary in interaction with the fare rules~\cite{taxirules2022}.

Both data sets were checked manually and iteratively for any errors. Conspicuous values were cleaned up. Cleaning was based on the error types described in Section~\ref{sec:data_errors}. In addition, rules were derived for subsequent error injection and for subsequent cleaning in streaming mode. The data sets were considered to be free of errors if all defined data quality rules are complied. To ensure that the data sets contained only controlled injected errors before starting the experiment, the data sets without any injected errors after the static cleaning were examined with the proposed framework. %outlier -> nicht in Paper

\subsubsection{Procedure}
The manually cleaned data sets were considered ground truth. Data errors were injected into this ground truth in a controlled manner. These were used in the context of the experiment to analyze error detection and repairing. Each of the error types described in Section~\ref{sec:data_errors} was examined separately. Interactions between different error types and their repairing were not considered.

For the prototypical evaluation, the data sets with the injected errors were first streamed through the different cleaning modules. The data set with injected errors as well as their positions and the positions where the detection algorithms found conspicuous values were stored in a database. The cleaned data set was also stored along with this information.

In order to evaluate the streaming pipeline's cleaning performance, seven metrics were defined, listed in Table~\ref{tab:metrics}.

\begin{table}
\centering
  \caption{Metrics for Evaluation of Cleaning Performance}
  \label{tab:metrics}
  \begin{tabular}{ p{0.3\linewidth}p{0.6\linewidth} }
    \toprule
    Metric & Description \\
    \midrule
     Identified errors & Number of vectors identified as errors by the pipeline\\
     Correctly identified errors & Number of vectors identified as errors, where an injected error was actually present\\
     False negatives & Number of vectors that were not identified as errors, even though an error was injected\\
     False positives & Number of vectors identified as errors, although no error was injected\\
     Number of deleted vectors & Number of vectors processed in the pipeline for which the error repairing was performed by rejecting the vector\\
     Mean value ratio & Ratio of the mean value of the column values of the ground truth to the mean value of the columns after repairing\\
     Standard deviation ratio & Ratio of the standard deviation of the column values of the ground truth to the mean value of the columns after repairing\\
  \bottomrule
\end{tabular}
\end{table}
The metrics are presented separately for each column of a data set and additionally for the entire data set (as far as this makes sense for the metric). The results tables with metrics for all cleaning runs for both data sets are shown in the appendix of this paper. They are discussed below.

\subsection{Results}
The results of the experiments are described in this section. It could be shown that:
\begin{itemize}
    \item The practical evaluation confirmed the theoretical findings
    \item Time dependency in particular has a strong influence on the results
\end{itemize}

The detailed results are discussed separately for each type of error below. The referenced results tables can be found in the appendix of this paper. Not all error types were meaningful for both data sets. Table~\ref{tab:cleaning-results} contains an overview of which error types were examined in which data sets as well as an overview of the results described in detail below. As mentioned, error type \emph{terminology heterogeneity} was excluded and error type \emph{wrong data type} is handled in advance. Therefore, these error types do not appear in this table.

\begin{table}
\centering
  \caption{Cleaning Results per Error Type}
  \label{tab:cleaning-results}
  \begin{tabular}{ p{0.2\linewidth} >{\centering\arraybackslash}p{0.075\linewidth} >{\centering\arraybackslash}p{0.075\linewidth} p{0.25\linewidth} p{0.25\linewidth} }
    \toprule
    Error Type & \multicolumn{2}{c}{Examined} & Detection & Repairing \\
               & Intel & Taxi                 &           &           \\
    \midrule
     Uniqueness violation & \checkmark & - & Correct number of errors, wrong positions & Distributions after repairing close to ground truth\\
     Interval violation & \checkmark & \checkmark & All errors identified correctly & Not evaluable\\
     FD violation & - & \checkmark & All errors identified correctly, but many false positives & High deviations of the distributions\\
     Missing values & \checkmark & \checkmark & All errors identified correctly & Dependent on method\\
     Contradicting records & \checkmark & \checkmark & Correct number of errors, wrong positions & Distributions after repairing close to ground truth\\
     Duplicates & \checkmark & \checkmark & Correct number of errors, wrong positions & Distributions after repairing close to ground truth\\
     Outlier & \checkmark & - & Many false positives, partly low number of identified errors & Repairing results in inconsistencies\\
  \bottomrule
  \end{tabular}
\end{table}

\subsubsection{Uniqueness violation}
This error type was analyzed only for the Intel Lab Data Set. The number of detected errors corresponds to the number of injected errors. However, the positions often do not coincide. This effect occurs as the streaming data set was cleaned serially, but the errors were injected in parallel. The injected uniqueness violation is equally likely to be temporally before or after the data vector whose uniqueness was violated. As described, due to time dependency, the first occurrence of the unique vector does not result in an error. Only the second occurrence leads to an identifiable error. 

\subsubsection{Interval violation}
Interval violations were examined in both data sets, the results can be seen in the appendix in Table~\ref{tab:intel-interval_violation} and Table~\ref{tab:taxi-interval_violation}. All injected errors were fully detected. The positions were also consistent. Thus, no false positives occurred.

In the experiment, a cost-based approach was used for repairing of this error type. This has the advantage that a replacement value can be generated from the values of the current vector alone. Access to other vectors is not necessary. Thus, repairing is not affected by time dependence. However, this also leads to the fact that the evaluability of the results is impaired. The comparison of mean and standard deviation of the adjusted data set with the corresponding ground truth values implicitly evaluates the performance of distribution-based repair approaches. Therefore, an assessment of the repairing performance is not reasonably possible with the existing metrics.

\subsubsection{Functional dependency violation}
This error type was analyzed for the New York Taxi Data Set only (see Table~\ref{tab:taxi-fd_violation} in the appendix). The results show high false positive rates. Still, all injected errors could be detected. This effect results from the interaction of several functional dependencies valid in parallel for one attribute. The errors injected for each functional dependency were correctly identified. However, unintentional violations of other functional dependencies may have occurred during error injection for one functional dependency that are not understood to be purposefully injected errors.

After repairing, high differences were found between the distribution measures of many attributes. This results from the fact that to inject an error, the value of attributes was changed. However, this changed value does not necessarily have to be corrected. The repairing can also be achieved by changing another value of attributes involved in the functional dependency. An approximation of the distribution measures of individual attributes to the values of the ground truth can therefore not be expected in the experiment.

\subsubsection{Missing values}
Missing values were examined in both data sets. The results can be found in the appendix in Table~\ref{tab:intel-missing_values} and Table~\ref{tab:taxi-missing_values}. In both cases, all injected errors were detected at the correct position. This confirms that detection of missing values is not affected by the specifics of data streams.

The results of the repairing are highly dependent on the chosen method. Different methods were analyzed for different attributes as part of the experiment. The results show that missing value imputation by statistical distribution measures is not fundamentally unsuitable in this experiment. However, it must be taken into account that these characteristic distribution measures may change with each new data vector. The concrete repair value therefore depends on the time of repairing. For data sets where the distribution varies greatly over time, imputation using distribution-based methods would not be appropriate.

\subsubsection{Contradicting records}
This type of error was analyzed in both data sets (see Table~\ref{tab:intel-contradicting_records} and Table~\ref{tab:taxi-contradicting_records} in the appendix). The non-uniqueness of key attributes was used as an indicator. For both data sets, the number of identified errors equals the number of injected errors and the number of vectors removed by repairing. However, the positions of the errors do not match. Only about half of the errors were identified at the location where they were injected. As described for error type \emph{uniqueness violation}, this is due to the time dependence in streaming pipelines, where the first vector is always evaluated as error-free.

\subsubsection{Duplicates}
The error type was investigated in both data sets, as can be seen in the appendix in Table~\ref{tab:intel-duplicates} and Table~\ref{tab:taxi-duplicates}. The results correspond to those of error type \emph{contradicting records}. Both types of errors are highly affected by time dependency. The order of data in the stream affects the detection of errors.

\subsubsection{Outlier}
Outliers were examined only for the Intel Lab Data Set and the results can be found in Table~\ref{tab:intel-outlier}. The number of correctly identified errors varied considerably depending on the attribute.  Overall, only 85.42\% of the injected defects were detected correctly. The fewest errors could be identified with 57.38\% at attribute \emph{light}. Most errors were identified with 99.41\% at attribute \emph{temperature}. However, 90 false positives were also identified for this attribute, which corresponds to 13.22\%.

The closest value that was not classified as an outlier was used for repairing. The dispersion of the attribute values was reduced by this method. However, in the results of the experiment, all defective attributes have significantly increased standard deviations after repairing. One possible explanation is that the chosen methods do not match the statistical properties of the attribute values. This emphasizes the importance of carefully selecting repairing methods that fit the application domain and the characteristics of the data set.

For handling of the error type \emph{outlier} in data streams no sufficient solution could be found in this work. The high complexity of this type of error, in the context of the particular properties of data streams identified in this work, provide a starting point for future research.

\paragraph{\textbf{Summary}}
The practical evaluation confirmed the theoretical findings. The time dependency in particular has a strong influence on the results. It is therefore recommended to avoid algorithms that are based on distribution values in a streaming context. In cases where such methods are not available, solutions must be found for dealing with time dependency. Either directly during data cleaning or during subsequent data use.

Particularly in highly automated applications such as data streams, a special potential for increasing data quality lies in preventing errors from occurring in the first place. Conceivable solutions that provide potential for in-depth investigation include redundancies, control and checking mechanisms during data collection, or the collection of additional context data.

\section{Technologies}
\label{sec:technologies}
The focus of the previous section was on the practical evaluation of the theoretical findings. For this reason, only the errors and their cleaning were discussed at this point. This section, on the other hand, analyzes the special requirements of data cleaning for data streaming technologies. Different data streaming technologies are compared.% The necessity of stateful streaming for the different error types is also discussed.

\subsection{Requirements}
First, the specific requirements for choosing the appropriate data streaming technology for data cleaning were analyzed. Requirement dimensions were derived from the general classification by researchers Kolajo et al. and Isah et al. by reflecting the characteristics of the relevant error types~\cite{Kolajo2019,Isah2019}. First of all, a data streaming technology should have a \emph{native streaming engine} to enable fast data cleaning. A native streaming engine is generally more advantageous than a micro-batch engine, because the incoming data stream does not need to be transformed into a batch. Secondly, the data streaming technology should offer an \emph{exactly-once guarantee} so that no additional errors are induced in the data cleaning pipeline by duplicating data streams. Moreover, it is important that the data streaming technology enables the developer to \emph{examine the elements of a data stream}. If errors are found, the technology needs to offer means to correct them by \emph{transformations}. For some error types, the data cleaning pipeline needs to store values for a certain time interval to compare new data vectors with older ones (\emph{stateful streaming}). As this is closely dependent on time dependency, it is discussed in more detail below.

\subsection{Stateful Streaming}
To implement the cleaning functions it is necessary, as described, for some error types to consider multiple data vectors at the same time. This is strongly related to the time dependency of the error types, as Table~\ref{tab:stateful-streaming} shows. Stateful streaming can be used to store relevant characteristics of already processed data vectors using user-defined states. For example, it may be relevant to store the unique identifiers of the processed data vectors of the last 60 seconds to identify any \emph{uniqueness violation}. New data vectors can then be compared to older ones. The application of windows on the data stream is also considered as a stateful operation. Windows are used to group data vectors and aggregate values. For some error types (such as detecting \emph{outlier} and imputing \emph{missing values}), windows are necessary to calculate statistical values like the local mean or standard deviation.

\begin{table}
\centering
  \caption{Need for Stateful Streaming -- \ding{55} indicates that the error type is affected by time dependency}
  \label{tab:stateful-streaming}
  \begin{tabular}{ p{0.25\linewidth} >{\centering\arraybackslash}p{0.2\linewidth} >{\centering\arraybackslash}p{0.15\linewidth}}
    \toprule
    Error Type & Cleaning Function  & Time dependent \\
    \midrule
     \emph{Schema Level} & & \\
     \hspace{3mm}Uniqueness violation & Stateful & \ding{55}\\
     \hspace{3mm}Wrong data type & Stateless & --\\
     \hspace{3mm}Interval violation & Stateless & --\\
     \hspace{3mm}FD violation & Stateless & --\\
     \emph{Instance Level} & & \\
     \hspace{3mm}Missing values & Stateful & \ding{55}\\
     \hspace{3mm}Duplicates & Stateful & \ding{55}\\
     \hspace{3mm}Outlier & Stateful & \ding{55}\\
     \hspace{3mm}Contradicting records & Stateful & \ding{55}\\
  \bottomrule
\end{tabular}
\end{table}

\subsection{Comparison of Data Streaming Technologies}
The comparison of data streaming technologies is discussed in the following based on the requirements presented. The literature reviews most often referred to the data streaming technologies Apache Storm, Apache Spark and Apache Flink~\cite{Kolajo2019,Isah2019}. Therefore, those technologies were selected to assess the fulfillment of the requirement dimensions for data cleaning. Table~\ref{tab:classification-results} shows the results of the comparison. They are described in more detail for each technology below.

\begin{table}
\centering
  \caption{Comparison of Data Streaming Technologies -- \checkmark indicates that the requirement is fulfilled}
  \label{tab:classification-results}
  \begin{tabular}{ p{0.275\linewidth} >{\centering\arraybackslash}p{0.15\linewidth} >{\centering\arraybackslash}p{0.15\linewidth} >{\centering\arraybackslash}p{0.15\linewidth} }
    \toprule
      & Apache Storm & Apache Spark & Apache Flink \\
    \midrule
     Native streaming engine & \checkmark & -- & \checkmark\\
     Exactly once guarantee & -- & \checkmark & \checkmark\\
     Examination mechanisms & \checkmark & \checkmark & \checkmark\\
     Transformations & \checkmark & \checkmark & \checkmark\\
     Stateful streaming & \checkmark & \checkmark & \checkmark\\
  \bottomrule
  \end{tabular}
\end{table}

\subsubsection{Apache Storm}
Apache Storm uses four abstractions to define data streaming applications: topology, data stream, spouts and bolts~\cite{Storm2023d}. A topology is a graph consisting of spouts and bolts and can be interpreted as the logical view of a Storm application. Data streams are defined in line with the definition in this paper. The spouts are the data stream sources~\cite{Storm2023d}. The bolts are used to process and transform the data streams. For complex transformations, multiple bolts might be necessary, each of them applying a simple transformation on the data stream~\cite{Storm2023a}.
Apache Storm offers a native streaming engine to process data streams. It also provides an at-least-once guarantee, which means that each data vector is processed at least once and might be replayed during failures. Moreover, data vectors can be examined and transformed using either bolts or base operations of the Apache Storm Stream API. For window-based operations, Apache Storm offers both sliding and tumbling windows~\cite{Storm2023g}. Bolts can store and query the state of their operations~\cite{Storm2023f,Isah2019}. Per default, the state of a bolt is saved every second~\cite{Storm2023f}. Apache Storm also provides the opportunity to define custom states as key-value pairs~\cite{Isah2019,Storm2023f}. In total, Apache Storm fulfills all requirements except for the exactly-once guarantee. The technology cannot guarantee that each data vector is processed exactly once.

\subsubsection{Apache Spark}
According to Khalid and Yousaf, the development of Apache Spark aimed at expanding on the MapReduce model by Hadoop~\cite{Khalid2021}. Therefore, Apache Spark is based on the abstraction of resilient distributed datasets~\cite{Spark2023c}. Resilient distributed datasets are partitions that can be distributed among multiple nodes in a Spark cluster~\cite{Khalid2021}. 
Apache Spark processes data streams with the Structured Streaming Engine that is based on the Spark SQL Engine~\cite{Spark2023b,Spark2023c}. The main idea behind the processing model is the interpretation of a data stream as an continuously growing table~\cite{Spark2023c}. The table is periodically updated to add incoming data vectors as new rows. The update frequency can be defined by the user~\cite{Spark2023c}. Secondly, Apache Spark uses offsets to provide an exactly-once guarantee for suitable data stream sources and sinks~\cite{Spark2023c}. A data stream can be examined and transformed by modelling queries on Datasets or DataFrames. Window-based operations are possible for tumbling, sliding and session windows~\cite{Spark2023b}. However, Apache Spark only defines windows in relation to time intervals and not with regards to the number of elements within a window~\cite{Spark2023b}. For stateful streaming, the technology provides state stores which can save key-value pairs over several batches~\cite{Spark2023b}. 
Summing up, Apache Spark fulfills all requirements except for the native streaming engine.

\subsubsection{Apache Flink}
Apache Flinks offers multiple abstraction levels to define data streaming applications~\cite{Flink2023c}. The DataStream API is one of the main APIs to Apache Flink and is an abstraction of data streaming applications on the level of higher programming languages. With the DataStreamAPI, a data streaming application can be created by calling functions in a streaming execution environment.
Apache Flink uses a native streaming engine to process data streams~\cite{Flink2023}. For suitable data stream sources and sinks the exactly-once guarantee can be applied~\cite{Flink2023e}. Data streams can be modeled with the Apache Flink data type DataStream. Tumbling, sliding, session and global windows can be used for window-based transformations~\cite{Flink2023h}. For the implementation of stateful streaming, Apache Flink offers so-called rich functions that are extensions to the basic stream operations. Moreover, user-defined states can be used in combination with values, lists or aggregated values~\cite{Flink2023g}. In order to use stateful operations the data stream needs to be grouped by a specified key~\cite{Flink2023g}.
In conclusion, Apache Flink is the only examined data streaming technology that fulfills all requirements. 

\section{Conclusion and Future Work}
\label{sec:conclusion}
In this paper, we conducted a fundamental investigation of data cleaning of data streams. The detection, repair and applicability to streams for relevant error types were systematically analyzed. We furthermore investigated how the special characteristics of data streams affect the cleaning of these error types. The theoretical analyses were evaluated practically using a prototype framework with different data sets. 

The results show that two particularities of streams affect data cleaning: the need for automated processing and time dependency. Data streams, unlike static data sets, must be cleaned immediately and made available for further applications. In addition, the data set is never fully known. Important distribution measures, which are often used for data cleaning, but also the uniqueness of attribute values are thus not constant over time. Therefore, two identical vectors can receive different error classifications at different times. Cleaning is not consistent in this case.

Further topics for future research emerge from the presented results. In the context of this work, only the cleaning of single error types in a pipeline was analyzed. The combination of multiple error types and their detection and repairing rules in the streaming context could help to improve the data cleaning quality. 

The effects of windows and windows sizes on data quality should also be examined more closely. In this work, it was shown that the time dependency already has an effect on the data quality if the entire stream $S_m$ and time $t_m$ is available. This is rarely the case in real streaming applications, which is why windows are used. It is thus to be expected that the choice of window and its size has a significant impact on data quality. Future studies should investigate how the optimal window type and size can be found for a use case.

Moreover, this work has focused on streams in general rather than time series in particular. For time series, there are further possibilities for cleaning, e.g. by interpolation. A detailed study of these possibilities has to be carried out.

Furthermore, data streams were distinguished from static data sets in this work. In practice, however, a data set is often extended incrementally in batches. The role of time dependency for such use cases has not yet been addressed and must be done accordingly.

\section*{Acknowledgement}
Many thanks to Meike Klettke for the helpful feedback.

\bibliographystyle{unsrt}  
\bibliography{bibliography}  %%% Remove comment to use the external .bib file (using bibtex).

\eject
\section*{Appendix}
\label{appendix}
\begin{table*}[h]
\caption{Uniqueness violation, Intel Lab Data Set}
\label{tab:intel-uniqueness_violation}
\begin{tabular}{l|rrrrrr}
	\toprule
	& \textbf{timestamp}   & \textbf{temperature}   & \textbf{humidity}   & \textbf{light}   & \textbf{voltage}   & \textbf{Total}   \\
	\midrule
	Number of errors     & 1008        & 0             & 0          & 0       & 0         & 1008     \\
	Identified errors & 1008        & 0             & 0          & 0       & 0         & 1008     \\
	Correctly identified errors    & 524         & 0             & 0          & 0       & 0         & 524      \\
    \% correctly identified errors  & 51.98 \%     & n/a           & n/a        & n/a     & n/a       & 51.98 \%  \\
	Unidentified errors      & 484         & 0             & 0          & 0       & 0         & 484      \\
	False positives          & 484         & 0             & 0          & 0       & 0         & 484      \\
	\% false positives         & 48.02 \%     & n/a           & n/a        & n/a     & n/a       & 48.02 \%  \\
	Number of deleted vectors               &             &               &            &         &           & 1008     \\
	Mean value ratio             & 100.00 \%      & 99.96 \%        & 99.74 \%     & 99.50 \%   & 100.04 \%   & 99.81 \%   \\
	Standard deviation ratio    & 99.95 \%      & 100.67 \%       & 100.57 \%    & 99.65 \%  & 100.45 \%   & 100.26 \%  \\
	\bottomrule
\end{tabular}
\end{table*}

\begin{table*}[h]
\caption{Interval violation, Intel Lab Data Set}
\label{tab:intel-interval_violation}
\begin{tabular}{l|rrrrrr}
	\toprule
	& \textbf{timestamp}   & \textbf{temperature}   & \textbf{humidity}   & \textbf{light}   & \textbf{voltage}   & \textbf{Total}   \\
	\midrule
	Number of errors     & 0           & 504           & 490        & 495     & 527       & 2016     \\
	Identified errors & 0           & 504           & 490        & 495     & 527       & 2016     \\
	Correctly identified errors    & 0           & 504           & 490        & 495     & 527       & 2016     \\
    \% correctly identified errors  & n/a         & 100.00 \%       & 100.00 \%    & 100.00 \% & 100.00 \%   & 100.00 \%  \\
	Unidentified errors      & 0           & 0             & 0          & 0       & 0         & 0        \\
	False positives          & 0           & 0             & 0          & 0       & 0         & 0        \\
	\% false positives         & n/a         & 0.00 \%         & 0.00 \%      & 0.00 \%   & 0.00 \%     & 0.00 \%    \\
	Number of deleted vectors               &             &               &            &         &           & 0        \\
	Mean value ratio             & 100.00 \%      & 100.34 \%       & 101.17 \%    & 97.59 \%  & 97.38 \%    & 99.12 \%   \\
	Standard deviation ratio    & 100.00 \%      & 99.39 \%        & 104.38 \%    & 99.91 \%  & 556.25 \%   & 191.99 \%  \\
	\bottomrule
\end{tabular}
\end{table*}

\begin{table*}
\caption{Missing values, Intel Lab Data Set}
\label{tab:intel-missing_values}
\begin{tabular}{l|rrrrrr}
	\toprule
	& \textbf{timestamp}   & \textbf{temperature}   & \textbf{humidity}   & \textbf{light}   & \textbf{voltage}   & \textbf{Total}   \\
	\midrule
	Number of errors     & 0           & 504           & 495        & 490     & 527       & 2016     \\
	Identified errors    & 0           & 504           & 495        & 490     & 527       & 2016     \\
	Correctly identified errors     & 0           & 504           & 495        & 490     & 527       & 2016     \\
    \% correctly identified errors  & n/a         & 100.00 \%       & 100.00 \%    & 100.00 \% & 100.00 \%   & 100.00 \%  \\
	Unidentified errors        & 0           & 0             & 0          & 0       & 0         & 0        \\
	False positives            & 0           & 0             & 0          & 0       & 0         & 0        \\
	\% false positives         & n/a         & 0.00 \%         & 0.00 \%      & 0.00 \%   & 0.00 \%     & 0.00 \%    \\
	Number of deleted vectors  &             &               &            &         &           & 0        \\
	Mean value ratio           & 100.00 \%      & 99.96 \%        & 99.78 \%     & 99.41 \%  & 100.03 \%   & 99.79 \%   \\
	Standard deviation ratio   & 100.00 \%      & 98.81 \%        & 99.22 \%     & 98.79 \%  & 98.94 \%    & 99.15 \%   \\
	\bottomrule
\end{tabular}
\end{table*}

\begin{table*}
\caption{Contradicting Records, Intel Lab Data Set}
\label{tab:intel-contradicting_records}
\begin{tabular}{l|rrrrrr}
	\toprule
	& \textbf{timestamp}   & \textbf{temperature}   & \textbf{humidity}   & \textbf{light}   & \textbf{voltage}   & \textbf{Total}   \\
	\midrule
	Number of errors     & 1008        & 0             & 0          & 0       & 0         & 1008     \\
	Identified errors & 1008        & 0             & 0          & 0       & 0         & 1008     \\
	Correctly identified errors    & 512         & 0             & 0          & 0       & 0         & 512      \\
    \% correctly identified errors  & 50.79 \%     & n/a           & n/a        & n/a     & n/a       & 50.79 \%  \\
	Unidentified errors      & 496         & 0             & 0          & 0       & 0         & 496      \\
	False positives          & 496         & 0             & 0          & 0       & 0         & 496      \\
	\% false positives         & 49.21 \%     & n/a           & n/a        & n/a     & n/a       & 49.21 \%  \\
	Number of deleted vectors               &             &               &            &         &           & 1008     \\
	Mean value ratio             & 100.00 \%      & 99.97 \%        & 99.75 \%     & 99.66 \%  & 100.04 \%   & 99.86 \%   \\
	Standard deviation ratio    & 99.87 \%      & 100.63 \%       & 100.57 \%    & 99.71 \%  & 100.43 \%   & 100.24 \%  \\
	\bottomrule
\end{tabular}
\end{table*}

\begin{table*}
\caption{Duplicates, Intel Lab Data Set}
\label{tab:intel-duplicates}
\begin{tabular}{l|rrrrrr}
	\toprule
	& \textbf{timestamp}   & \textbf{temperature}   & \textbf{humidity}   & \textbf{light}   & \textbf{voltage}   & \textbf{Total}   \\
	\midrule
	Number of errors     & 1008        & 1008          & 1008       & 1008    & 1008      & 5040     \\
	Identified errors    & 1008        & 1008          & 1008       & 1008    & 1008      & 5040     \\
	Correctly identified errors     & 524         & 524           & 524        & 524     & 524       & 2620     \\
    \% correctly identified errors  & 51.98 \%     & 51.98 \%       & 51.98 \%    & 51.98 \% & 51.98 \%   & 51.98 \%  \\
	Unidentified errors        & 484         & 484           & 484        & 484     & 484       & 2420     \\
	False positives            & 484         & 484           & 484        & 484     & 484       & 2420     \\
	\% false positives         & 48.02 \%     & 48.02 \%       & 48.02 \%    & 48.02 \% & 48.02 \%   & 48.02 \%  \\
	Number of deleted vectors  &             &               &            &         &           & 1008        \\
	Mean value ratio           & 100.00 \%      & 99.96 \%        & 100.02 \%    & 99.96 \%  & 100.00 \%    & 99.98 \%   \\
	Standard deviation ratio   & 99.99 \%      & 100.08 \%       & 99.92 \%     & 99.88 \%  & 99.86 \%    & 99.95 \%   \\
	\bottomrule
\end{tabular}
\end{table*}

\begin{table*}
\caption{Outlier, Intel Lab Data Set}
\label{tab:intel-outlier}
\begin{tabular}{l|rrrrrr}
	\toprule
	& \textbf{timestamp}   & \textbf{temperature}   & \textbf{humidity}   & \textbf{light}   & \textbf{voltage}   & \textbf{Gesamt}   \\
	\midrule
	Number of errors     & 0           & 681           & 671        & 664     & 0         & 2016     \\
	Identified errors & 0           & 767           & 664        & 381     & 0         & 1812     \\
	Correctly identified errors    & 0           & 677           & 664        & 381     & 0         & 1722     \\
 	\% correctly identified errors  & n/a         & 99.41 \%       & 98.96 \%    & 57.38 \% & n/a       & 85.42 \%  \\
	Unidentified errors      & 0           & 4             & 7          & 283     & 0         & 294      \\
	False positives          & 0           & 90            & 0          & 0       & 0         & 90       \\
	\% false positives         & n/a         & 13.22 \%       & 0.0 \%      & 0.0 \%   & n/a       & 4.46 \%   \\
	Number of deleted vectors               &             &               &            &         &           & 0        \\
	Mean value ratio             & 100.0\%      & 97.48\%        & 97.2\%      & 97.04\%  & 100.0\%    & 97.93\%   \\
	Standard deviation ratio    & 100.0\%      & 124.47\%       & 121.83\%    & 107.12\% & 100.0\%    & 110.68\%  \\
	\bottomrule
\end{tabular}
\end{table*}

%\subsection{New York Taxi Data Set}
\begin{table*}
\caption{Interval violation, New York Taxi Data Set}
\label{tab:taxi-interval_violation}
\begin{tabular}{l|rrrrrrrrrr}
	\toprule
	
	                       & \rotatebox{90}{Number of errors} & \rotatebox{90}{Identified errors} & \rotatebox{90}{Correctly identified errors} & \rotatebox{90}{\% correctly identified errors} & \rotatebox{90}{Unidentified errors} & \rotatebox{90}{False positives} & \rotatebox{90}{\% false positives} & \rotatebox{90}{Number of deleted vectors} & \rotatebox{90}{Mean value ratio} & \rotatebox{90}{Standard deviation ratio} \\
\midrule
tpep\_pickup\_time     & 1389                         & 1389                        & 1389                     & 0                      & 0                       & 100.00 \%                    & 0.0 \%            &                    & 100.01 \%             & 6510.96 \%                     \\
tpep\_dropoff\_time    & 0                            & 0                           & 0                        & 0                      & 0                       & n/a                         & n/a               &                    & 100.00 \%              & 100.00 \%                       \\
passenger\_count       & 1396                         & 1396                        & 1396                     & 0                      & 0                       & 100.00 \%                    & 0.0 \%            &                    & 102.48 \%             & 104.03 \%                      \\
trip\_distance         & 2700                         & 2688                        & 2688                     & 12                     & 0                       & 99.56 \%                    & 0.0 \%            &                    & 97.52 \%              & 99.42 \%                       \\
RatecodeID             & 1366                         & 1366                        & 1366                     & 0                      & 0                       & 100.00 \%                    & 0.0 \%            &                    & n/a                  & n/a                           \\
PULocationID           & 1326                         & 1326                        & 1326                     & 0                      & 0                       & 100.00 \%                    & 0.0 \%            &                    & n/a                  & n/a                           \\
DOLocationID           & 1360                         & 1360                        & 1360                     & 0                      & 0                       & 100.00 \%                    & 0.0 \%            &                    & n/a                  & n/a                           \\
payment\_type          & 1343                         & 1343                        & 1343                     & 0                      & 0                       & 100.00 \%                    & 0.0 \%            &                    & n/a                  & n/a                           \\
fare\_amount           & 0                            & 0                           & 0                        & 0                      & 0                       & n/a                         & n/a               &                    & 100.00 \%              & 100.00 \%                       \\
extra                  & 0                            & 0                           & 0                        & 0                      & 0                       & n/a                         & n/a               &                    & 100.00 \%              & 100.00 \%                       \\
mta\_tax               & 0                            & 0                           & 0                        & 0                      & 0                       & n/a                         & n/a               &                    & 100.00 \%              & 100.00 \%                       \\
tip\_amount            & 0                            & 0                           & 0                        & 0                      & 0                       & n/a                         & n/a               &                    & 100.00 \%              & 100.00 \%                       \\
tolls\_amount          & 0                            & 0                           & 0                        & 0                      & 0                       & n/a                         & n/a               &                    & 100.00 \%              & 100.00 \%                       \\
improvement\_surcharge & 0                            & 0                           & 0                        & 0                      & 0                       & n/a                         & n/a               &                    & 100.00 \%              & 100.00 \%                       \\
total\_amount          & 0                            & 0                           & 0                        & 0                      & 0                       & n/a                         & n/a               &                    & 100.00 \%              & 100.00 \%                       \\
congestion\_surcharge  & 0                            & 0                           & 0                        & 0                      & 0                       & n/a                         & n/a               &                    & 100.00 \%              & 100.00 \%                       \\
airport\_fee           & 0                            & 0                           & 0                        & 0                      & 0                       & n/a                         & n/a               &                    & 100.00 \%              & 100.00 \%                       \\
RideID                 & 0                            & 0                           & 0                        & 0                      & 0                       & n/a                         & n/a               &                    & 100.00 \%              & 100.00 \%                       \\
timestamp              & 0                            & 0                           & 0                        & 0                      & 0                       & n/a                         & n/a               &                    & 100.00 \%              & 100.00 \%                       \\
storno\_flag           & 0                            & 0                           & 0                        & 0                      & 0                       & n/a                         & n/a               &                    & n/a                  & n/a                           \\
Total                  & 10880                        & 10868                       & 10868                    & 12                     & 0                       & 99.89 \%                    & 0.0 \%            & 0                  & 100.00 \%              & 527.63\%                      \\

\bottomrule
\end{tabular}
\end{table*}

\begin{table*}
\caption{Functional dependency violation, New York Taxi Data Set}
\label{tab:taxi-fd_violation}
\begin{tabular}{l|rrrrrrrrrr}
	\toprule
	
	                       & \rotatebox{90}{Number of errors} & \rotatebox{90}{Identified errors} & \rotatebox{90}{Correctly identified errors} & \rotatebox{90}{\% correctly identified errors} & \rotatebox{90}{Unidentified errors} & \rotatebox{90}{False positives} & \rotatebox{90}{\% false positives} & \rotatebox{90}{Number of deleted vectors} & \rotatebox{90}{Mean value ratio} & \rotatebox{90}{Standard deviation ratio} \\
\midrule
tpep\_pickup\_time     & 2640                         & 2640                        & 2640                     & 0                      & 0                       & 100.00 \%                    & 0.00 \%            &                    & 100.00 \%              & 100.00 \%                       \\
tpep\_dropoff\_time    & 2640                         & 2640                        & 2640                     & 0                      & 0                       & 100.00 \%                    & 0.00 \%            &                    & 100.00 \%              & 100.00 \%                       \\
passenger\_count       & 0                            & 0                           & 0                        & 0                      & 0                       & n/a                         & n/a               &                    & 100.00 \%              & 100.00 \%                       \\
trip\_distance         & 2701                         & 3656                        & 2701                     & 0                      & 955                     & 100.00 \%                    & 35.36 \%          &                    & 107.60 \%              & 103.57 \%                      \\
RatecodeID             & 5515                         & 5515                        & 5515                     & 0                      & 0                       & 100.00 \%                    & 0.00 \%            &                    & n/a                  & n/a                           \\
PULocationID           & 2701                         & 3656                        & 2701                     & 0                      & 955                     & 100.00 \%                    & 35.36 \%          &                    & n/a                  & n/a                           \\
DOLocationID           & 0                            & 0                           & 0                        & 0                      & 0                       & n/a                         & n/a               &                    & n/a                  & n/a                           \\
payment\_type          & 0                            & 0                           & 0                        & 0                      & 0                       & n/a                         & n/a               &                    & n/a                  & n/a                           \\
fare\_amount           & 5551                         & 8189                        & 5551                     & 0                      & 2638                    & 100.00 \%                    & 47.52 \%          &                    & 116.92 \%             & 256.85 \%                      \\
extra                  & 2737                         & 5551                        & 2737                     & 0                      & 2814                    & 100.00 \%                    & 102.81 \%         &                    & 117.00 \%              & 290.98 \%                      \\
mta\_tax               & 2737                         & 5551                        & 2737                     & 0                      & 2814                    & 100.00 \%                    & 102.81 \%         &                    & 106.81 \%             & 1018.14 \%                     \\
tip\_amount            & 2737                         & 5551                        & 2737                     & 0                      & 2814                    & 100.00 \%                    & 102.81 \%         &                    & 108.28 \%             & 245.84 \%                      \\
tolls\_amount          & 2737                         & 5551                        & 2737                     & 0                      & 2814                    & 100.00 \%                    & 102.81 \%         &                    & 122.26 \%             & 182.38 \%                      \\
improvement\_surcharge & 2737                         & 5551                        & 2737                     & 0                      & 2814                    & 100.00 \%                    & 102.81 \%         &                    & 107.68 \%             & 1175.87 \%                     \\
total\_amount          & 2737                         & 5551                        & 2737                     & 0                      & 2814                    & 100.00 \%                    & 102.81 \%         &                    & 109.75 \%             & 215.06 \%                      \\
congestion\_surcharge  & 2737                         & 5551                        & 2737                     & 0                      & 2814                    & 100.00 \%                    & 102.81 \%         &                    & 107.07 \%             & 534.06 \%                      \\
airport\_fee           & 2737                         & 5551                        & 2737                     & 0                      & 2814                    & 100.00 \%                    & 102.81 \%         &                    & 182.28 \%             & 498.02 \%                      \\
RideID                 & 0                            & 0                           & 0                        & 0                      & 0                       & n/a                         & n/a               &                    & 100.00 \%              & 100.00 \%                       \\
timestamp              & 0                            & 0                           & 0                        & 0                      & 0                       & n/a                         & n/a               &                    & 100.00 \%              & 100.00 \%                       \\
storno\_flag           & 0                            & 0                           & 0                        & 0                      & 0                       & n/a                         & n/a               &                    & n/a                  & n/a                           \\
Total                  & 43644                        & 70704                       & 43644                    & 0                      & 27060                   & 100.00 \%                    & 62.00 \%           &                    & 115.47 \%             & 334.72 \%                      \\
\bottomrule
\end{tabular}
\end{table*}

\begin{table*}
\caption{Missing values, New York Taxi Data Set}
\label{tab:taxi-missing_values}
\begin{tabular}{l|rrrrrrrrrr}
	\toprule
	
	                       & \rotatebox{90}{Number of errors} & \rotatebox{90}{Identified errors} & \rotatebox{90}{Correctly identified errors} & \rotatebox{90}{\% correctly identified errors} & \rotatebox{90}{Unidentified errors} & \rotatebox{90}{False positives} & \rotatebox{90}{\% false positives} & \rotatebox{90}{Number of deleted vectors} & \rotatebox{90}{Mean value ratio} & \rotatebox{90}{Standard deviation ratio} \\
\midrule
tpep\_pickup\_time     & 0                            & 0                           & 0                        & 0                      & 0                       & n/a                         & n/a               &                    & 100.00 \%              & 100.00 \%                       \\
tpep\_dropoff\_time    & 0                            & 0                           & 0                        & 0                      & 0                       & n/a                         & n/a               &                    & 100.00 \%              & 100.00 \%                       \\
passenger\_count       & 732                          & 732                         & 732                      & 0                      & 0                       & 100.00 \%                    & 0.0 \%            &                    & 100.01 \%             & 99.74 \%                       \\
trip\_distance         & 766                          & 766                         & 766                      & 0                      & 0                       & 100.00 \%                    & 0.0 \%            &                    & 100.00 \%              & 99.66 \%                       \\
RatecodeID             & 727                          & 727                         & 727                      & 0                      & 0                       & 100.00 \%                    & 0.0 \%            &                    & n/a                  & n/a                           \\
PULocationID           & 740                          & 740                         & 740                      & 0                      & 0                       & 100.00 \%                    & 0.0 \%            &                    & n/a                  & n/a                           \\
DOLocationID           & 735                          & 735                         & 735                      & 0                      & 0                       & 100.00 \%                    & 0.0 \%            &                    & n/a                  & n/a                           \\
payment\_type          & 704                          & 704                         & 704                      & 0                      & 0                       & 100.00 \%                    & 0.0 \%            &                    & n/a                  & n/a                           \\
fare\_amount           & 759                          & 759                         & 759                      & 0                      & 0                       & 100.00 \%                    & 0.0 \%            &                    & 99.99 \%              & 99.63 \%                       \\
extra                  & 742                          & 742                         & 742                      & 0                      & 0                       & 100.00 \%                    & 0.0 \%            &                    & 99.65 \%              & 99.80 \%                        \\
mta\_tax               & 693                          & 693                         & 693                      & 0                      & 0                       & 100.00 \%                    & 0.0 \%            &                    & 100.01 \%             & 99.77 \%                       \\
tip\_amount            & 703                          & 703                         & 703                      & 0                      & 0                       & 100.00 \%                    & 0.0 \%            &                    & 99.98 \%              & 99.75 \%                       \\
tolls\_amount          & 742                          & 742                         & 742                      & 0                      & 0                       & 100.00 \%                    & 0.0 \%            &                    & 99.67 \%              & 99.50 \%                        \\
improvement\_surcharge & 688                          & 688                         & 688                      & 0                      & 0                       & 100.00 \%                    & 0.0 \%            &                    & 100.00 \%              & 100.00 \%                       \\
total\_amount          & 758                          & 758                         & 758                      & 0                      & 0                       & 100.00 \%                    & 0.0 \%            &                    & 100.00 \%              & 99.62 \%                       \\
congestion\_surcharge  & 728                          & 728                         & 728                      & 0                      & 0                       & 100.00 \%                    & 0.0 \%            &                    & 99.98 \%              & 99.71 \%                       \\
airport\_fee           & 675                          & 675                         & 675                      & 0                      & 0                       & 100.00 \%                    & 0.0 \%            &                    & 99.98 \%              & 99.67 \%                       \\
RideID                 & 0                            & 0                           & 0                        & 0                      & 0                       & n/a                         & n/a               &                    & 100.00 \%              & 100.00 \%                       \\
timestamp              & 0                            & 0                           & 0                        & 0                      & 0                       & n/a                         & n/a               &                    & 100.00 \%              & 100.00 \%                       \\
storno\_flag           & 0                            & 0                           & 0                        & 0                      & 0                       & n/a                         & n/a               &                    & n/a                  & n/a                           \\
Total                  & 10892                        & 10892                       & 10892                    & 0                      & 0                       & 100.00 \%                    & 0.0 \%            & 0                  & 99.94 \%              & 99.79 \%                       \\
\bottomrule
\end{tabular}
\end{table*}

\begin{table*}
\caption{Contradicting Records, New York Taxi Data Set}
\label{tab:taxi-contradicting_records}
\begin{tabular}{l|rrrrrrrrrr}
	\toprule
	
	                       & \rotatebox{90}{Number of errors} & \rotatebox{90}{Identified errors} & \rotatebox{90}{Correctly identified errors} & \rotatebox{90}{\% correctly identified errors} & \rotatebox{90}{Unidentified errors} & \rotatebox{90}{False positives} & \rotatebox{90}{\% false positives} & \rotatebox{90}{Number of deleted vectors} & \rotatebox{90}{Mean value ratio} & \rotatebox{90}{Standard deviation ratio} \\
\midrule
tpep\_pickup\_time     & 0                & 0                 & 0                           & n/a                            & 0                   & 0               & n/a                &                           & 100.00\%          & 100.25\%                \\
tpep\_dropoff\_time    & 0                & 0                 & 0                           & n/a                            & 0                   & 0               & n/a                &                           & 100.00\%          & 100.21\%                \\
passenger\_count       & 0                & 0                 & 0                           & n/a                            & 0                   & 0               & n/a                &                           & 100.00\%          & 100.05\%                \\
trip\_distance         & 0                & 0                 & 0                           & n/a                            & 0                   & 0               & n/a                &                           & 99.80\%           & 100.00\%                \\
RatecodeID             & 0                & 0                 & 0                           & n/a                            & 0                   & 0               & n/a                &                           & n/a              & n/a                      \\
PULocationID           & 0                & 0                 & 0                           & n/a                            & 0                   & 0               & n/a                &                           & n/a              & n/a                      \\
DOLocationID           & 0                & 0                 & 0                           & n/a                            & 0                   & 0               & n/a                &                           & n/a              & n/a                      \\
payment\_type          & 0                & 0                 & 0                           & n/a                            & 0                   & 0               & n/a                &                           & n/a              & n/a                      \\
fare\_amount           & 0                & 0                 & 0                           & n/a                            & 0                   & 0               & n/a                &                           & 99.94\%          & 100.00\%                 \\
extra                  & 0                & 0                 & 0                           & n/a                            & 0                   & 0               & n/a                &                           & 97.87\%          & 99.48\%                  \\
mta\_tax               & 0                & 0                 & 0                           & n/a                            & 0                   & 0               & n/a                &                           & 100.02\%         & 99.46\%                  \\
tip\_amount            & 0                & 0                 & 0                           & n/a                            & 0                   & 0               & n/a                &                           & 99.82\%          & 99.91\%                  \\
tolls\_amount          & 0                & 0                 & 0                           & n/a                            & 0                   & 0               & n/a                &                           & 99.97\%          & 99.70\%                  \\
improvement\_surcharge & 0                & 0                 & 0                           & n/a                            & 0                   & 0               & n/a                &                           & 100.01\%         & 99.47\%                  \\
total\_amount          & 0                & 0                 & 0                           & n/a                            & 0                   & 0               & n/a                &                           & 99.89\%          & 99.92\%                  \\
congestion\_surcharge  & 0                & 0                 & 0                           & n/a                            & 0                   & 0               & n/a                &                           & 100.01\%         & 99.88\%                  \\
airport\_fee           & 0                & 0                 & 0                           & n/a                            & 0                   & 0               & n/a                &                           & 99.57\%          & 99.83\%                  \\
RideID                 & 5446             & 5446              & 2824                        & 51.85 \%                       & 2622                & 2622            & 48.15 \%           &                           & 100.02\%         & 99.99\%                  \\
timestamp              & 0                & 0                 & 0                           & n/a                            & 0                   & 0               & n/a                &                           & 100.00\%          & 100.21\%                \\
storno\_flag           & 5446             & 5446              & 2824                        & 51.85 \%                       & 2622                & 2622            & 48.15 \%           &                           & n/a              & n/a                      \\
Total                  & 10892            & 10892             & 5648                        & 51.85 \%                       & 5244                & 5244            & 48.15 \%           &  5446                     & 99.74\%          & 99.89\%  \\                
\bottomrule
\end{tabular}
\end{table*}

\begin{table*}
\caption{Duplicates, New York Taxi Data Set}
\label{tab:taxi-duplicates}
\begin{tabular}{l|rrrrrrrrrr}
	\toprule
	
	                       & \rotatebox{90}{Number of errors} & \rotatebox{90}{Identified errors} & \rotatebox{90}{Correctly identified errors} & \rotatebox{90}{\% correctly identified errors} & \rotatebox{90}{Unidentified errors} & \rotatebox{90}{False positives} & \rotatebox{90}{\% false positives} & \rotatebox{90}{Number of deleted vectors} & \rotatebox{90}{Mean value ratio} & \rotatebox{90}{Standard deviation ratio} \\
\midrule
tpep\_pickup\_time     & 5446                         & 5446                        & 2715                     & 2731                   & 2731                    & 49.85 \%                    & 50.15 \%          &                    & 100.00 \%              & 99.99 \%                       \\
tpep\_dropoff\_time    & 5446                         & 5446                        & 2715                     & 2731                   & 2731                    & 49.85 \%                    & 50.15 \%          &                    & 100.00 \%              & 100.00 \%                       \\
passenger\_count       & 5446                         & 5446                        & 2715                     & 2731                   & 2731                    & 49.85 \%                    & 50.15 \%          &                    & 100.00 \%              & 99.94 \%                       \\
trip\_distance         & 5446                         & 5446                        & 2715                     & 2731                   & 2731                    & 49.85 \%                    & 50.15 \%          &                    & 99.98 \%              & 99.96 \%                       \\
RatecodeID             & 5446                         & 5446                        & 2715                     & 2731                   & 2731                    & 49.85 \%                    & 50.15 \%          &                    & n/a                  & n/a                           \\
PULocationID           & 5446                         & 5446                     & 2715                     & 2731                   & 2731                       & 49.85 \%                    & 50.15 \%          &                & n/a                  & n/a                           \\
DOLocationID           & 5446                         & 5446                        & 2715                     & 2731                   & 2731                    & 49.85 \%                    & 50.15 \%          &                    & n/a                  & n/a                           \\
payment\_type          & 5446                         & 5446                        & 2715                     & 2731                   & 2731                    & 49.85 \%                    & 50.15 \%          &                    & n/a                  & n/a                           \\
fare\_amount           & 5446                         & 5446                        & 2715                     & 2731                   & 2731                    & 49.85 \%                    & 50.15 \%          &                    & 99.97 \%              & 99.94 \%                       \\
extra                  & 5446                         & 5446                        & 2715                     & 2731                   & 2731                    & 49.85 \%                    & 50.15 \%          &                    & 100.06\%             & 100.06 \%                      \\
mta\_tax               & 5446                         & 5446                        & 2715                     & 2731                   & 2731                    & 49.85 \%                    & 50.15 \%          &                    & 100.01 \%             & 99.86 \%                       \\
tip\_amount            & 5446                         & 5446                        & 2715                     & 2731                   & 2731                    & 49.85 \%                    & 50.15 \%          &                    & 99.96 \%              & 99.95 \%                       \\
tolls\_amount          & 5446                         & 5446                        & 2715                     & 2731                   & 2731                    & 49.85 \%                    & 50.15 \%          &                    & 99.79 \%              & 99.77 \%                       \\
improvement\_surcharge & 5446                         & 5446                        & 2715                     & 2731                   & 2731                    & 49.85 \%                    & 50.15 \%          &                    & 100.00 \%              & 100.10 \%                       \\
total\_amount          & 5446                         & 5446                        & 2715                     & 2731                   & 2731                    & 49.85 \%                    & 50.15 \%          &                    & 99.98 \%              & 99.96 \%                       \\
congestion\_surcharge  & 5446                         & 5446                        & 2715                     & 2731                   & 2731                    & 49.85 \%                    & 50.15 \%          &                    & 100.01 \%             & 100.04 \%                      \\
airport\_fee           & 5446                         & 5446                        & 2715                     & 2731                   & 2731                    & 49.85 \%                    & 50.15 \%          &                    & 10026 \%             & 100,10 \%                       \\
RideID                 & 5446                         & 5446                        & 2715                     & 2731                   & 2731                    & 49.85 \%                    & 50.15 \%          &                    & 100.03 \%             & 99,97 \%                       \\
timestamp              & 5446                         & 5446                        & 2715                     & 2731                   & 2731                    & 49.85 \%                    & 50.15 \%          &                    & 100.00 \%              & 100.00 \%                       \\
storno\_flag           & 5446                         & 5446                        & 2715                     & 2731                   & 2731                    & 49.85 \%                    & 50.15 \%          &                    & n/a                  & n/a                           \\
Total                  & 108920                       & 108920                      & 54300                    & 54620  & 54620   & 49.85 \%  & 50.15 \%   & 5446   & 100.00 \%   & 99.98 \% \\
\bottomrule
\end{tabular}
\end{table*}

\eject

\end{document}